\documentclass[twocolumn]{article}
\usepackage{mystyle,epsf,amssymb,amsmath}

\begin{document}

\twocolumn[
\vspace*{30mm}
\centerline{\LARGE Universal level-spacing statistics in quasiperiodic 
tight-binding models}\vspace{3ex}
\centerline{\large Uwe Grimm$^{1}$, 
                   Rudolf A.\ R\"{o}mer$^{1}$, 
                   Michael Schreiber$^{1}$,
                   Jian-Xin Zhong$^{2,3}$}\vspace{2ex}
\begin{footnotesize}
\centerline{
\begin{minipage}{0.75\textwidth}
${}^{1}${\it Institut f\"{u}r Physik, 
             Technische Universit\"{a}t, 
             D--09107 Chemnitz, Germany}\\
${}^{2}${\it Department of Physics, University of Tennessee, 
             Knoxville, Tennessee 37996, USA; and}\\ 
\hphantom{${}^{2}$}{\it
             Solid State Division,
             Oak Ridge National Laboratory,
             Oak Ridge, Tennessee 37831--6032, USA}\\
${}^{3}${\it Department of Physics, Xiangtan University,
             Xiangtan 411105, P.~R.~China}\\
\end{minipage}
}\end{footnotesize}
\centerline{\footnotesize August 4, 1999}\vspace{4ex}
\begin{small}
\hrule\vspace{2ex}
\begin{minipage}{\textwidth}
{\bf Abstract}\vspace{2ex}\\
\hp 
We study statistical properties of the energy spectra of
two-dimensional quasiperiodic tight-binding models.  The multifractal
nature of the eigenstates of these models is corroborated by the
scaling of the participation numbers with the systems size.  Hence one
might have expected ``critical'' or ``intermediate'' statistics for
the level-spacing distributions as observed at the metal-insulator
transition in the three-dimensional Anderson model of
disorder. However, our numerical results are in perfect agreement with
the universal level-spacing distributions of the Gaussian orthogonal
random matrix ensemble, including the distribution of spacings between
second, third, and forth neighbour energy levels.\vspace{2ex}
\end{minipage}\vspace{2ex}
\hrule
\end{small}\vspace{6ex}
]

\section{Introduction}

\hp The anomalous transport properties of quasicrystals evince the
peculiar electronic properties of these intermetallic alloys, which
are reflected in the nature and the density of electronic states in
quasiperiodic systems. Tight-binding models on quasiperiodic tilings
serve as simple toy models of electrons moving in a quasiperiodic
environment. While much is known rigorously for one-dimensional
aperiodic Schr\"{o}dinger operators \cite{Suto}, quasiperiodic
tight-binding models in the physically relevant dimensions two and
three have mainly been investigated numerically, compare \cite{RGS}
and references therein for the few exceptions, which, however, yield
only rather limited information about the overall structure of the
spectrum or the wave functions.

At least in two dimensions, numerical results provide ample evidence
that typical eigenstates of quasiperiodic tight-binding Hamiltonians
are, as in the one-dimensional case, neither extended nor
exponentially localized; instead, the probability amplitudes show a
multifractal distribution \cite{RS}. However, the integrated density
of states does not show the hierarchical plateau structure observed
for one-dimensional substitution chains \cite{BBG}; for pure hopping
Hamiltonians it appears to be rather smooth, apart from a few gaps in
the spectrum. This makes it possible to employ spectral statistics
\cite{mehta} as further tools to characterize the energy spectrum of
quasiperiodic tight-binding models.

\begin{figure}[t]
\centerline{\epsfxsize=0.8\columnwidth\epsfbox{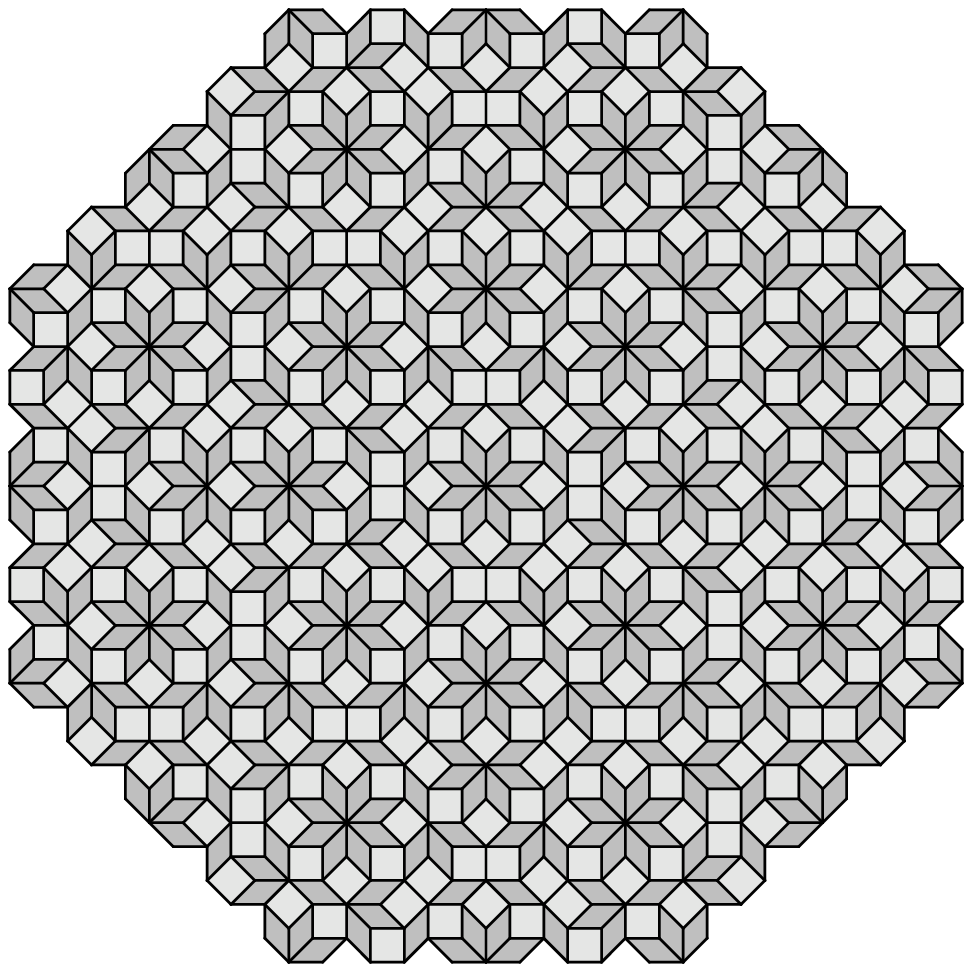}\vspace{1ex}}
\centerline{\epsfxsize=0.6\columnwidth\epsfbox{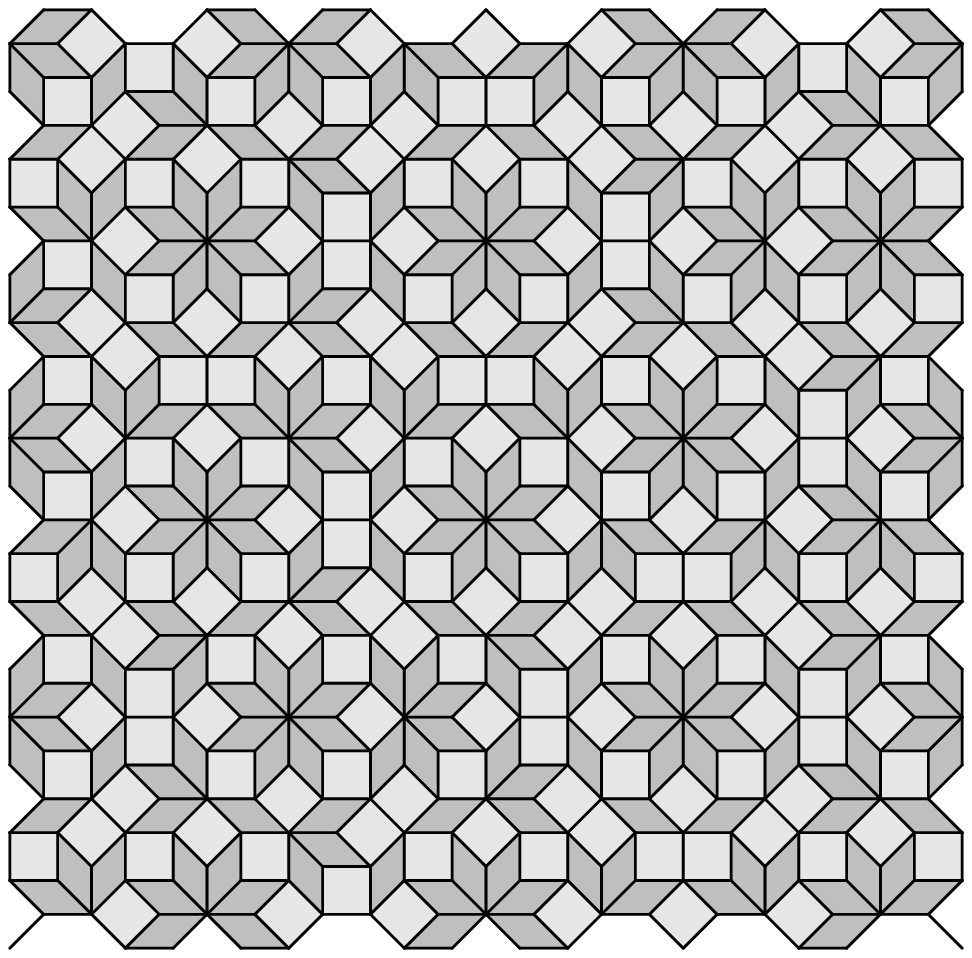}\vspace{-1ex}}
\caption{\label{fig-tiling}
    Top: Octagonal cluster of the Ammann-Beenker tiling with 833
    vertices and exact $D_8$-symmetry around the central vertex.
    Bottom: A square-shaped cut with 496 vertices without any exact 
    spatial symmetries.}
\end{figure}

\section{Energy-level statistics}

\hp 
The statistical analysis of the spectra of quantum systems has a rich
history, and is at the center stage of current interest \cite{S}. For
disordered systems, it is by now well known that in the metallic
regime the level-spacing distribution $P_0(s)$ of neighbouring
eigenenergies is well described by $P^{\rm GOE}_0(s)$ of the Gaussian
orthogonal ensemble (GOE) of random matrix theory \cite{mehta}. Here,
$s$ denotes the energy spacing in units of the mean level spacing, and
$P_0(s)$ gives the distribution of normalized gaps in the spectrum in
the limit of infinite system size. In the insulating regime, the
level-spacing distribution follows the Poisson law $P^{\rm P}_0(s)=
\exp(-s)$. At the metal-insulator transition, the level-spacing
distribution has been shown to follow yet another behavior which is
attributed to the existence of a ``critical ensemble''
\cite{SSSLS,HS,ZK97}.

\begin{figure}[t]
\centerline{\epsfxsize=\columnwidth\epsfbox{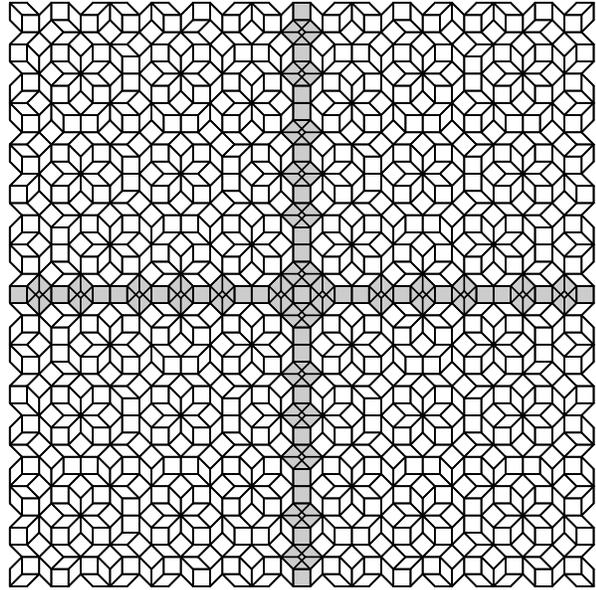}\vspace{-1ex}}
\caption{\label{fig-approx} Periodic approximant of the Ammann-Beenker tiling 
with 1393 vertices, displayed together with a copy of itself that has been
rotated through 90 degrees. The ``worms'' where mismatches between the two 
rotated copies occur are shaded.}
\end{figure}

Recently, we have shown that the level statistics of tight-binding
models defined on planar quasiperiodic tilings, notably the
Ammann-Beenker tiling shown in Fig.~\ref{fig-tiling}, is also given by
the GOE predictions \cite{QPLSD1}, contrary to previous statements in
the literature \cite{BS,PJ,ZY}. The deviations from GOE behaviour
observed in \cite{BS,PJ} could be understood by noting that the
standard periodic approximants of the octagonal tiling are singular
patches whose fourfold rotational symmetry is broken only weakly, see
Fig.~\ref{fig-approx}. This ``almost symmetry'' of the patch does not
lead to a block structure of the Hamiltonian (which would mean
independent spectra for each block), but nevertheless affects the
level-spacing distribution. As shown in \cite{QPLSD1}, the
level-spacing distribution $P_0(s)$ of generic patches of the tiling,
or of the irreducible sectors of patches with exact symmetries,
follows the GOE distribution when applying a suitable unfolding
procedure to correct for the fluctuations in the density of states.
This result was further corroborated by higher correlations such as
the Mehta-Dyson statistics $\Delta_3$ \cite{QPLSD1} and the number
variance $\Sigma_2$ \cite{QPLSD2}.  Furthermore, we have shown that
the level-spacing distribution remains GOE if one considers small
parts of the spectrum without any unfolding, and we demonstrated that
the unfolding procedure has to be performed somewhat differently from
the procedure in disordered systems \cite{QPLSD3}.

In this paper, we numerically calculate not only the first, but also
the second, the third, and the fourth neighbour level-spacing
distributions $P_n(s)$, $n=0, \ldots, 3$, respectively. We show that
the higher level-spacing distributions are also well described by the
predictions of random matrix theory. The recently proposed
``semi-Poisson ensemble'' \cite{BG}, supposed to be valid for the
multifractal eigenstates of the metal-insulator transition in the
three-dimensional Anderson model \cite{braun}, does not describe the
higher level-spacing distributions of the quasiperiodic models. The
same applies to other intermediate statistics obtained recently for an
ensemble constructed by an analogy with a one-dimensional gas model of
particles with nearest neighbour interactions $V(x)= -\log|x|$
\cite{gerland}. This is intermediate between the Poisson and the GOE
statistics. which can also be interpreted in terms of Boltzmann
distributions of one-dimensional interacting gases with pair
potentials $V(x)=0$ and $V(x)= -\log|x|$, but {\em not}\/ restricted
to nearest neighbours, respectively.  Finally, we calculate the
participation numbers in these tilings and show that their scaling
behavior indicates multifractal behavior.

\section{Higher level-spacing distributions}

\hp
In addition to the nearest-neighbour level-spacing distribution
$P_0(s)$ considered previously \cite{QPLSD1,QPLSD2,QPLSD3}, we have
now computed the distributions $P_1(s)$, $P_2(s)$, and $P_3(s)$ for
the simple Hamiltonian 
\begin{equation}
H = \sum_{ij}^{N} t_{ij} |i\rangle \langle j|
\label{ham}
\end{equation}
defined on the Ammann-Beenker or octagonal tiling as shown in
Fig.~\ref{fig-tiling}. Here, each vertex $i=1,2,\ldots,N$ of the
tiling carries a state $|i\rangle$, and the hopping elements
$t_{ij}=t_{ji}$ are chosen to be unity between neighbouring vertices,
i.e.\ vertices connected by a bond, and zero otherwise. The spectral
statistics are obtained by numerically diagonalizing the Hamiltonian
(\ref{ham}) with free boundary conditions for large patches of the
tiling, taking into account their symmetries.  While the GOE
predictions for these distributions are known in principle
\cite{mehta}, it is non-trivial to obtain numerical data for these
functions. As we could not find sufficiently accurate results in the
literature, we calculated $P_0^{\rm GOE}(s)$, $P_1^{\rm GOE}(s)$,
$P_2^{\rm GOE}(s)$, and $P_3^{\rm GOE}(s)$ using the approach of
\cite{tracy} by numerically solving the appropriate differential
equations. This was done using {\em Mathematica}\/ by performing a
small-$s$ expansion, followed by a numerical integration to larger
values of $s$. The small-$s$ behaviours of the GOE distributions are
$P_0^{\rm GOE}(s) = \pi^2 s/6 + {\cal O}(s^3)$, $P_1^{\rm GOE}(s) =
\pi^4 s^4/270 + {\cal O}(s^6)$, $P_2^{\rm GOE}(s) = \pi^8 s^8/1764000
+ {\cal O}(s^{10}) $, and $P_3^{\rm GOE}(s) = \pi^{12}
s^{13}/168781725000 + {\cal O}(s^{15})$. In our calculations we
evaluated the expansions beyond the 30th order.

\begin{figure}[t]
\centerline{\epsfxsize=\columnwidth\epsfbox{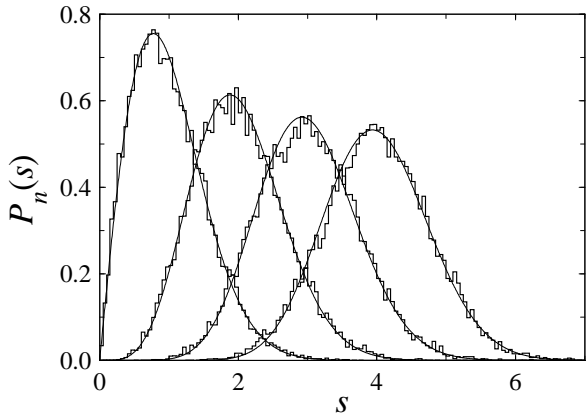}\vspace{-1ex}}
\caption{\label{fig-PS-1234-GOE}
    Level-spacing distribution up to the forth neighbour spacing for the 
    largest sector of the octagonal patch with $N=157369$ vertices. 
    The thin lines represent the GOE predictions \protect\cite{tracy}.}
\end{figure}

\begin{figure}[t]
\centerline{\epsfxsize=\columnwidth\epsfbox{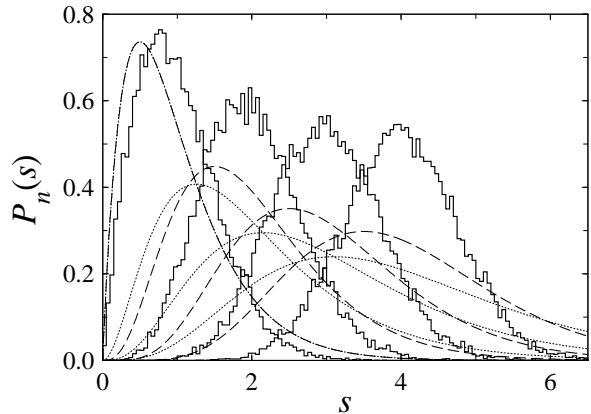}\vspace*{-1ex}}
\caption{\label{fig-PS-1234-SP-NNG} 
    Comparison of the level-spacing distributions of 
    Fig.~\protect\ref{fig-PS-1234-GOE}
    with the results for the semi-Poisson (dotted) and the
    one-dimensional gas model (dashed) ensembles. Note
    that $P_0(s)$ is identical for these two intermediate ensembles.}
\end{figure}

In Fig.~\ref{fig-PS-1234-GOE}, we show the results for the data of the
largest patch considered in \cite{QPLSD1}, an octagonal patch as in
Fig.~\ref{fig-tiling} but with $157369$ vertices, employing the
unfolding procedure described in \cite{QPLSD3}.  Apparently, our data
agree very well with the GOE distributions. We like to emphasize that
the higher level-spacing distributions considered here, and in
particular their small-$s$ behaviour, are very sensitive to detect
deviations from the GOE behaviour. However, we do not find deviations
beyond the fluctuations expected for finite systems. Thus, the
level-spacing distribution is the same as that of the Anderson model
of localization in the metallic regime \cite{SSSLS,HS}. This is
surprising, because it is well-known that eigenstates in the metallic
regime are extended, whereas there is ample evidence that the
eigenstates on planar quasiperiodic tilings have a multifractal
character. Thus one might speculate that the level statistics should
have been similar to the intermediate statistics found \cite{braun} at
the metal-insulator transition in three dimensions, where eigenstates
are multifractal as well. In Fig.~\ref{fig-PS-1234-SP-NNG}, we compare
our results to the level-spacing distributions of the intermediate
semi-Poisson ensemble \cite{BG,gerland} and the one-dimensional gas
ensemble with logarithmic interactions \cite{gerland}. These latter
statistics clearly do not agree with our data.

\section{Scaling of the participation numbers}

\begin{figure}[t]
\centerline{\epsfxsize=0.9\columnwidth\epsfbox{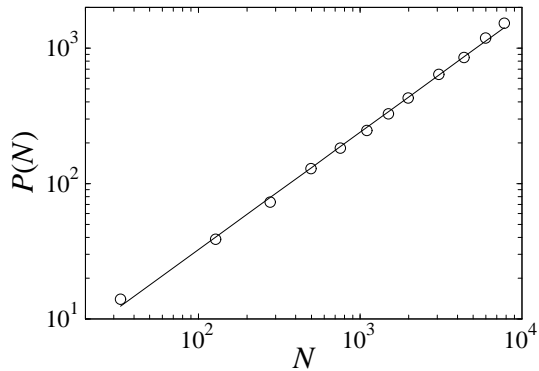}\vspace{-1ex}}
\caption{\label{fig-partnum} 
    Energy-averaged participation numbers 
    versus system size for square-shaped patches 
    (cp.\ Fig.\ \protect\ref{fig-tiling}). The straight line is a 
    least-squares fit of $\log P(N) = \kappa \log N + c$ to the data.}
\end{figure}

\hp 
We now turn our attention to the calculation of the participation
numbers $P(N,E)$ in order to investigate the properties of the states
in the tiling directly.  For this purpose, we consider square-shaped
patches as in Fig.~\ref{fig-tiling} with up to $N=7785$ vertices
\cite{QPLSD1} and compute
\begin{equation}
 P(N,E)^{-1}= \sum_{i=1}^{N} |\psi_i^4(E)|
\end{equation}
where $\psi(E)= (\psi_1(E), \ldots, \psi_N(E))$ denotes a normalized
eigenstate at energy $E$. Thus a state $\psi(E)$ that is completely
localized at site $j$, i.e.\ $\psi_i(E)= \delta_{i,j}$, has $P(N,E) =
1$, whereas a fully extended state $\psi_i(E)= 1/\sqrt{N}$ gives
$P(N,E) = N$. Because for the models under consideration the
level-spacing distribution does not appear to depend on the energy
\cite{QPLSD3}, we concentrate on the energy-averaged participation
number $P(N)$. For multifractal states, we expect to find a power-law
behaviour $P(N)\sim N^{\kappa}$ as a function of systems size
\cite{eilmes}, between the result for localized ($\kappa=0$) and
extended ($\kappa=1$) states. The result is shown in
Fig.~\ref{fig-partnum}, which yields a value of $\kappa \approx
0.87\pm0.05$. This value, in agreement with previous results for
quasiperiodic tight-binding models \cite{RS}, is clearly below the
values $\kappa=1$ for extended states, albeit rather close,
corroborating the multifractal character of the eigenstates in our
model, though with rather large characteristic exponents \cite{RS}.

\section{Conclusions}

\hp We considered the spectral properties of a tight-binding
Hamiltonian defined on the Ammann-Beenker tiling. Although this system
is not energetically disordered, the level-spacing distributions
exhibit level repulsion and are in perfect agreement with the GOE
predictions. This is similar as for the energetically disordered
Anderson model of localization in the metallic regime \cite{HS}, and
also for tight-binding models defined on certain random graphs
\cite{gudrun}.  Nevertheless, the scaling of the energy-averaged
participation numbers confirms that the corresponding eigenstates are
not extended. Their multifractal character is further corroborated by
the system-size dependence of the participation numbers.  At first
glance, it appears surprising that multifractal states may give rise
to GOE level statistics. However, it is known from studies of the
metal-insulator transition in more than two dimensions that the
critical level statistics becomes similar to the Poissonian behaviour
for larger dimensions \cite{ZK98}. Thus, it appears possible that the
critical level statistics for the two-dimensional case considered here
just coincides with the Gaussian orthogonal random matrix
ensemble. This scenario is in perfect agreement with the results
presented here.

\section*{Acknowledgements}

\hp
We thank C.\ A.\ Tracy for pointing out reference \cite{tracy} and for
suggesting how to compute $P^{\rm GOE}_n(s)$.  We also thank U.\ Gerland and
E.\ B.\ Bogomolny for stimulating discussions on intermediate
statistics.

\vspace{3ex}
\begin{footnotesize}

\end{footnotesize}

\end{document}